\documentclass[11pt]{article}

\usepackage{fontspec}
\defaultfontfeatures{Mapping=tex-text}
\usepackage{xunicode}
\usepackage{xltxtra}

\usepackage[a4paper, margin=2cm]{geometry}

\usepackage[UTF8]{ctex}
\usepackage[english]{babel}

\usepackage{amsmath, amssymb, bm}
\usepackage{slashed}

\usepackage{graphicx}
\usepackage{float}
\usepackage{subcaption}
\usepackage{booktabs}
\usepackage{siunitx}
\usepackage{multirow}

\usepackage{titlesec}
\titleformat{\section}
{\normalfont\Large\bfseries\centering}{\Roman{section}.}{1em}{}

\usepackage{cite}
\usepackage{indentfirst}
\usepackage{appendix}
\usepackage{diagbox}
\begin{document}

\begin{center}
	{\LARGE \textbf{
            $\mathcal{O}(v^4)$ corrections and resummation of relativistic corrections to gluon fragmentation into ${}^1S_0^{[1,8]}$ states
	}} \\[12pt]
	

  \large 
  Sai Cui$^{(a)}$, Guang-Zhi Xu$^{(a)}$, Yi-Jie Li$^{(a),*}$ and Kui-Yong Liu$^{(b,a),\dagger}$
  
  \vspace{6pt} 
  
  \normalsize 
  $^{(a)}$School of Physics, Liaoning University, Shenyang 110036, China\\
  $^{(b)}$School of Physics and Electronic Technology, Liaoning Normal University, Dalian 116029, China
  
  \vspace{6pt} 
  
  $*$Corresponding author. Email: liyijie@lnu.edu.cn\\
  $\dagger$Corresponding author. Email: liukuiyong@lnu.edu.cn
  
	{\small \today}
\end{center}

\begin{abstract}

Within the nonrelativistic QCD (NRQCD) factorization framework and based on the Collins--Soper operator definition of fragmentation functions, we compute the relativistic corrections to gluon fragmentation into pseudoscalar \(S\)-wave heavy-quark pairs in the
\({}^1S_0^{[1,8]}\)
channels.
By matching the full QCD partonic fragmentation function onto the NRQCD factorization formula, we extract the fixed-order short-distance coefficients through \(\mathcal O(v^4)\), building on the known leading-order  and \(\mathcal O(v^2)\) results.
We further perform the resummation of these relativistic corrections to obtain the fragmentation function.
Our results provide higher-order relativistic inputs for gluon fragmentation into pseudoscalar \(S\)-wave quarkonia and can be applied to precision predictions for
\(\eta_c\) and \(\eta_b\)
production in the large-transverse-momentum region.
	
\end{abstract}

\section{Introduction}

Heavy-quarkonium production provides an important probe of the interplay between perturbative and nonperturbative dynamics in quantum chromodynamics (QCD)
\cite{Gong:2008ft,Ma:2008gq,Zhang:2009ym,He:2009uf,Xu:2012am,Li:2013csa,Xu:2014zra}.
Since the heavy-quark mass \(m_Q\) is much larger than \(\Lambda_{\rm QCD}\), it defines a hard scale that can be treated within perturbative QCD, whereas quarkonium formation involves lower-energy scales associated with the relative momentum and binding energy of the heavy-quark pair.
This multiscale structure makes heavy quarkonium a useful testing ground for QCD factorization and effective field theory methods
\cite{Brambilla:2010cs}.

In the large-transverse-momentum region, \(p_T\gg m_Q\), the leading-power contribution to heavy-quarkonium production is dominated by the single-parton fragmentation mechanism
\cite{Kang:2011mg}.
The corresponding cross section can be written as a convolution of the partonic production cross section and the fragmentation function,
\begin{equation}
	d\sigma[A+B\to H+X]
	=
	\sum_i
	d\hat{\sigma}[A+B\to i+X]
	\otimes
	D_{i\to H}(z,\mu) .
	\label{eq:collinear-factorization}
\end{equation}
Here \(D_{i\to H}(z,\mu)\) denotes the fragmentation function for a parton \(i\) to fragment into a quarkonium state \(H\).
Collins and Soper introduced a gauge-invariant operator definition of fragmentation functions
\cite{Collins:1981uw,Collins1989}.
This definition expresses the fragmentation function as a nonlocal operator matrix element involving gauge links and provides the field-theoretical basis for calculating partonic fragmentation functions in full QCD.

Within the nonrelativistic quantum chromodynamics (NRQCD) factorization framework, the fragmentation function in Eq.~\eqref{eq:collinear-factorization} can be further expressed as a sum of products of short-distance coefficients (SDCs) and NRQCD long-distance matrix elements (LDMEs),
\begin{equation}
	D_{i\to H}(z,\mu)
	=
	\sum_n
	d_{i\to Q\bar{Q}(n)}(z,\mu)
	\langle {\cal O}^H_n\rangle .
	\label{eq:nrqcd-factorization-fragmentation}
\end{equation}
Here \(d_{i\to Q\bar{Q}(n)}(z,\mu)\) is a perturbatively calculable SDC that describes the short-distance production of a heavy-quark pair in the quantum state \(n\).
The corresponding LDME
\(\langle {\cal O}^H_n\rangle\)
describes the nonperturbative transition of the heavy-quark pair into the physical quarkonium state \(H\).
The SDCs are expanded in powers of the strong coupling constant \(\alpha_s\), while the LDMEs are organized according to the NRQCD velocity-scaling rules.
The fragmentation function can therefore be systematically expanded in both \(\alpha_s\) and the heavy-quark relative velocity \(v\).
Relativistic corrections correspond to higher-order terms in the velocity expansion and can be obtained by expanding the full QCD partonic fragmentation function and matching it order by order onto Eq.~\eqref{eq:nrqcd-factorization-fragmentation}
\cite{CaswellLepage1986,BBL1995}.

At leading order (LO) in both \(\alpha_s\) and \(v\), fragmentation functions for heavy-quark and gluon fragmentation into quarkonium states with different orbital angular momenta have been calculated systematically, including \(S\)-wave, \(P\)-wave, and \(D\)-wave channels
\cite{Ma:1994zt,Braaten:1993mp,Braaten:1993rw,Braaten:1994kd,Yuan:1994hn,Ma:1995ci,Ma:1995vi,Cheung:1995ir,Bodwin:2014bia}.
These results provide the basic matching benchmarks for calculations of QCD radiative corrections and relativistic corrections.
QCD radiative corrections have subsequently been calculated for several heavy-quark and gluon fragmentation channels at next-to-leading order (NLO) and beyond
\cite{Zheng:2019dfk,Zheng:2021ylc,Feng:2021uct,Zhang:2018mlo,Feng:2018ulg,Zhang:2020atv,Feng:2026owq}.
These studies demonstrate that higher-order QCD corrections can significantly affect the normalization, the \(z\) dependence, and the factorization-scale dependence of fragmentation functions and are therefore essential for precision predictions of quarkonium production at large transverse momentum.

Relativistic corrections constitute another important source of higher-order effects in quarkonium fragmentation functions.
The \(\mathcal O(v^2)\) corrections have been calculated for several \(S\)-wave and \(P\)-wave channels in heavy-quark and gluon fragmentation
\cite{Sang:2009zz,Cui:2026bwd,Bodwin:2003wh,Gao:2016ihc,He:2026nzm}.
By contrast, results at \(\mathcal O(v^4)\) remain limited.
For heavy-quark fragmentation, the \(\mathcal O(v^4)\) corrections to the color-singlet
\({}^1S_0^{[1]}\) and \({}^3S_1^{[1]}\)
channels have been calculated
\cite{Cui:2025wjq}.
For gluon fragmentation, the corresponding calculation has mainly been performed for the color-singlet
\({}^3S_1^{[1]}\)
channel
\cite{Bodwin:2012xc}.
An infrared (IR) divergence appears in this channel at \(\mathcal O(v^4)\), although no such divergence is present at lower orders.
The IR divergences cancel against poles in higher‑order perturbative color‑octet NRQCD LDMEs during matching.
This result shows that new IR structures may emerge at higher orders in the velocity expansion and that different spin channels must be examined independently.

Resummation of the relativistic corrections provides a complementary method for estimating higher-order contributions in the velocity expansion.
This method is commonly based on the generalized Gremm--Kapustin relation, which relates a class of higher-order relativistic LDMEs to lower-order nonperturbative inputs
\cite{Gremm:1997dq}.
Such resummation has been applied to heavy-quarkonium electromagnetic decays, double-quarkonium production, quarkonium-plus-photon production, and
\(e^+e^-\to H+\gamma\)
at a \(Z\) factory
\cite{Bodwin:2007fz,Chung:2010vz,Bodwin:2007ga,Fan:2012dy,Sang:2011fw,Lee:2010ts,Trunin:2024ibx}.
These studies show that resummation provides a useful means of estimating relativistic corrections beyond a fixed-order truncation.
However, previous applications have mainly focused on decay and exclusive production processes, while this resummation for fragmentation functions has not yet been studied systematically.

For gluon fragmentation into the color-singlet pseudoscalar \(S\)-wave heavy-quark pair
\({}^1S_0^{[1]}\),
the relativistic correction is currently known through \(\mathcal O(v^2)\)
\cite{Gao:2016ihc}.
The corresponding \(\mathcal O(v^2)\) SDC is proportional to the LO SDC, with a proportionality factor independent of the momentum fraction \(z\).
The NLO($\alpha_s$) radiative corrections for both the color‑singlet  ${}^1S_0^{[1]}$ and color‑octet ${}^1S_0^{[8]}$ channels have been obtained in Ref.~\cite{Feng:2018ulg} within the LO velocity expansion. Furthermore, the $\mathcal O(v^4)$ heavy‑quark‑fragmentation calculation for ${}^1S_0^{[1]}$ exhibits a sign difference relative to the $\mathcal O(v^2)$ result, offering insight into the convergence properties of the velocity expansion \cite{Cui:2025wjq}. 
In this respect, an \(\mathcal O(v^4)\) calculation for gluon fragmentation could test for similar possible sign variations and shed light on the velocity expansion relevant to \(\eta_{c,b}\) production.

Based on the known LO fragmentation function
\cite{Braaten:1993rw}
and the \(\mathcal O(v^2)\) result
\cite{Gao:2016ihc},
we calculate the \(\mathcal O(v^4)\) relativistic corrections to gluon fragmentation into the pseudoscalar \(S\)-wave heavy-quark pair
\({}^1S_0^{[1,8]}\).
Using the Collins--Soper definition of fragmentation functions and NRQCD matching, we derive the \(\mathcal O(v^4)\) SDC for the color-singlet
\({}^1S_0^{[1]}\)
channel and obtain the corresponding result for the color-octet
\({}^1S_0^{[8]}\)
channel.
We further calculate the resummed relativistic correction from the same finite-relative-momentum matching function by employing the generalized Gremm--Kapustin relation.
In the numerical analysis, we compare the LO result, the \(\mathcal O(v^2)\) and \(\mathcal O(v^4)\) corrections, and the resummed result, and examine the convergence of the velocity expansion and the numerical importance of the higher-order finite-relative-momentum corrections.

The rest of this paper is organized as follows.
In Sec.~\ref{sec:cs-nrqcd}, we introduce the Collins--Soper definition of the gluon fragmentation function and the NRQCD factorization formula.
In Sec.~\ref{sec:v4-correction}, we present the fixed-order calculation through \(\mathcal O(v^4)\) for gluon fragmentation into the color-singlet
\({}^1S_0^{[1]}\)
channel.
In Sec.~\ref{sec:relativistic-resummation}, we present the calculation of the resummed relativistic correction.
In Sec.~\ref{sec:numerical-results}, we give the numerical results for the
\({}^1S_0^{[1,8]}\)
channels and discuss the fixed-order corrections, the resummed results, and the convergence of the velocity expansion.
In Sec.~\ref{sec:summary}, we summarize our results.

\section{Collins--Soper Definition and NRQCD Formalism}
\label{sec:cs-nrqcd}

In this section, we establish the theoretical framework for gluon fragmentation into a color-singlet
\({}^1S_0^{[1]}\)
heavy-quark pair.
We begin with the gauge-invariant Collins--Soper definition of the gluon fragmentation function and extract the SDCs by matching the full QCD partonic fragmentation function onto its NRQCD representation.

We use the light-cone coordinates
\begin{equation}
	p^\mu=(p^+,p^-,\mathbf p_T),
	\qquad
	p^\pm=\frac{p^0\pm p^3}{\sqrt{2}},
	\qquad
	p\cdot q=p^+q^-+p^-q^+-\mathbf p_T\cdot\mathbf q_T .
	\label{eq:light-cone-coordinate}
\end{equation}
Here \(p^\pm\) are the light-cone components of the four-momentum \(p^\mu\), and \(\mathbf p_T\) is the transverse momentum.
The auxiliary lightlike vector is chosen as
\begin{equation}
	\hat n^\mu=(0,1,\mathbf 0_T),
	\qquad
	\hat n^2=0,
	\qquad
	\hat n\cdot A=A^+ .
	\label{eq:light-cone-vector}
\end{equation}
Here \(\hat n^\mu\) specifies the direction of the gauge link, and \(A^\mu\) is the gluon field.
With these conventions, the Collins--Soper definition of the gluon fragmentation function into a hadron \(H\) is
\begin{align}
	D_{g\to H}(z)
	=&
	\frac{-g_{\mu\nu}z^{d-3}}
	{2\pi k^+(N_c^2-1)(d-2)}
	\int_{-\infty}^{+\infty} dx^-\,
	e^{-ik^+x^-}
	\nonumber\\
	&\times
	\sum_X
	\langle 0|
	G_a^{+\mu}(0)\,
	\mathcal L[0,\infty]_{ab}
	|H(P^+,P^-,\mathbf P_T)+X\rangle
	\nonumber\\
	&\times
	\langle H(P^+,P^-,\mathbf P_T)+X|
	\mathcal L[\infty,x^-]_{bc}\,
	G_c^{+\nu}(0,x^-,\mathbf 0_T)
	|0\rangle .
	\label{eq:CS-gluon-common}
\end{align}
Here \(z=P^+/k^+\) is the longitudinal momentum fraction carried by \(H\), \(k^+\) is the plus component of the parent-gluon momentum, \(d\) is the spacetime dimension, and \(N_c\) is the number of colors.
The indices \(a\), \(b\), and \(c\) are adjoint color indices, \(G_a^{+\mu}\) is the gluon field-strength tensor, and \(\sum_X\) denotes the sum over all unobserved final states.
The factor \((N_c^2-1)(d-2)\) averages over the color and physical transverse polarizations of the parent gluon.

The light-cone momentum of the hadron \(H\) satisfies
\begin{equation}
	P^+=zk^+,
	\qquad
	P^-=\frac{M_H^2+\mathbf P_T^2}{2zk^+}.
	\label{eq:H-light-cone-momentum}
\end{equation}
Here \(M_H\) and \(\mathbf P_T\) are the mass and transverse momentum of \(H\), respectively.
The gauge link in Eq.~\eqref{eq:CS-gluon-common} is defined as
\begin{equation}
	\mathcal L[x^-,y^-]_{ab}
	=
	\left[
	\mathcal P
	\exp\left(
	-i g_s \int_{x^-}^{y^-} d\xi^-\,
	\hat n\cdot A(0,\xi^-,\mathbf 0_T)
	\right)
	\right]_{ab}.
	\label{eq:gauge-link-common}
\end{equation}
Here \(\mathcal P\) denotes path ordering, \(g_s\) is the QCD coupling, and
\(A^\mu=A_c^\mu T^c\)
is the matrix-valued gluon field in the adjoint representation.
The gauge link ensures the gauge invariance of the fragmentation function.

Within NRQCD factorization, the gluon fragmentation function into a heavy quarkonium \(H\) can be expressed as
\begin{equation}
	D_{g\to H}(z)
	=
	\sum_n
	d_{g\to Q\bar Q(n)}(z)\,
	\langle{\cal O}^H_n\rangle .
	\label{eq:NRQCD-factorization-gluon-general}
\end{equation}
Here \(n\) specifies the color, spin, and orbital angular momentum quantum numbers of the intermediate heavy-quark pair \(Q\bar Q(n)\).
The SDC
\(d_{g\to Q\bar Q(n)}(z)\)
describes the short-distance production of the heavy-quark pair and is determined through perturbative matching.
The corresponding LDME
\(\langle{\cal O}^H_n\rangle\)
describes the nonperturbative transition of the pair into the physical quarkonium state \(H\).

In the perturbative QCD calculations,
the heavy-quark and antiquark momenta are parametrized as
\begin{equation}
	p_Q=\frac{P}{2}+q,
	\qquad
	p_{\bar{Q}}=\frac{P}{2}-q,
	\qquad
	P\cdot q=0,
	\qquad
	P^2=4E^2,
	\qquad
	E=\sqrt{m^2+\mathbf q^2}.
	\label{eq:kinematics-common}
\end{equation}
Here \(P=p_Q+p_{\bar{Q}}\) is the total momentum of the heavy-quark pair,
\(q=(p_Q - p_{\bar{Q}})/2\)
is the relative momentum, \(m\) denotes the heavy-quark mass, and \(E\) is the energy of each heavy quark.

In the rest frame of the \(Q\bar Q\) pair,
\begin{equation}
	P=(2E,\mathbf 0),
	\qquad
	q=(0,\mathbf q),
	\qquad
	q^2=-\mathbf q^2 .
	\label{eq:rest-frame-kinematics}
\end{equation}
With nonrelativistic normalization, the perturbative NRQCD LDMEs for the
\(Q\bar Q({}^1S_0^{[1]})\)
state satisfy
\begin{equation}
	\langle {\cal O}^{Q\bar Q}_1({}^1S_0)\rangle
	=
	2N_c,
	\qquad
	\langle {\cal O}^{Q\bar Q}_{1,2r}({}^1S_0)\rangle
	=
	2N_c\,(\mathbf q^2)^r .
	\label{eq:perturbative-LDME-general}
\end{equation}
Here \(r\) is a nonnegative integer, and
\(\langle {\cal O}^{Q\bar Q}_{1,2r}({}^1S_0)\rangle\)
denotes the perturbative color-singlet LDME containing \(2r\) spatial covariant derivatives.

The partonic fragmentation function for gluon fragmentation into the free heavy-quark pair can therefore be written as
\begin{equation}
	D_{g\to Q\bar Q({}^1S_0^{[1]})}(z,\mathbf q^2)
	=
	2N_c
	\sum_{r=0}^{\infty}
	d^{(2r)}(z)
	\left(\frac{\mathbf q^2}{m^2}\right)^r .
	\label{eq:partonic-expanded-general}
\end{equation}
Eq.~\eqref{eq:partonic-expanded-general} relates the perturbative QCD fragmentation function to its NRQCD velocity expansion. 
\(d^{(2r)}(z)\) denotes the SDC at \(\mathcal O(v^{2r})\).

\begin{figure}[htbp]
	\centering
	\includegraphics[width=0.45\textwidth]{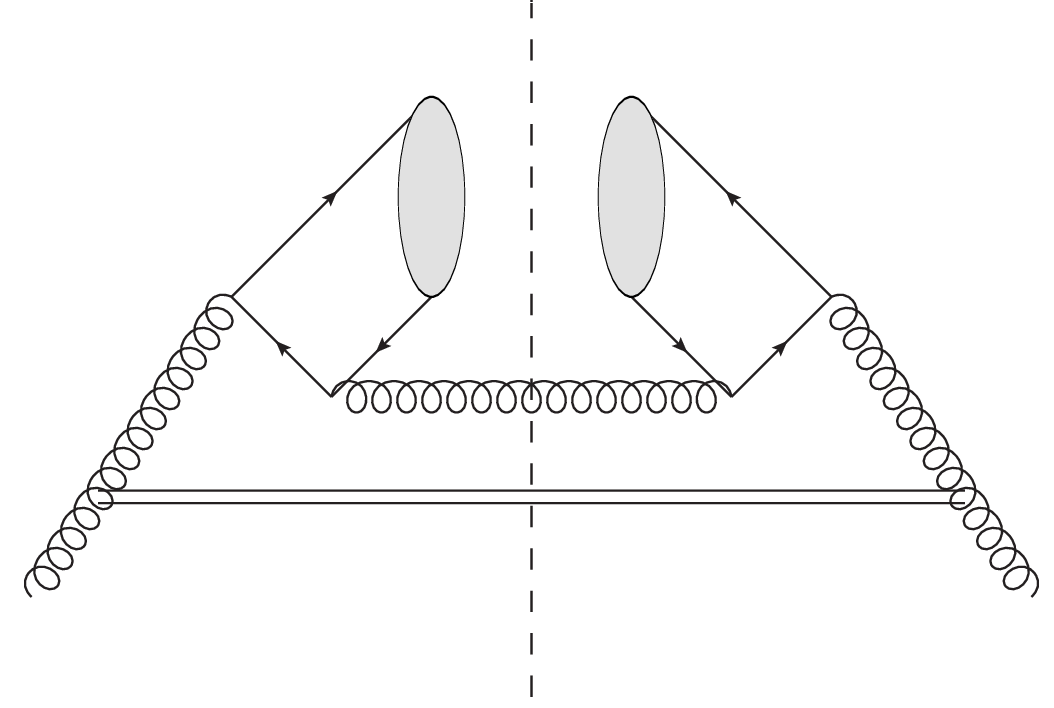}
	\caption{
		LO cut diagrams for gluon fragmentation into a color-singlet
		\({}^1S_0^{[1]}\)
		heavy-quark pair.
	}
	\label{fig0}
\end{figure}

At LO, the partonic fragmentation function for gluon fragmentation into the color-singlet
\({}^1S_0^{[1]}\)
heavy-quark pair is obtained from the cut diagrams shown in Fig.~\ref{fig0}.

According to the Collins--Soper definition, the corresponding partonic fragmentation function is
\begin{align}
	D_{g\to Q\bar Q({}^1S_0^{[1]})}(z,\mathbf q^2)
	=&
	\frac{\alpha_s^2 E}{N_c^2-1}
	\int
	\frac{dl^+\,d^2\mathbf l_T}{(2\pi)^3 2l^+}
	\int
	\frac{dP^+\,d^2\mathbf P_T}{(2\pi)^3 2P^+}
	\nonumber\\
	&\times
	2\pi\delta(zk^+-P^+)
	\delta(k^+-P^+-l^+)
	\delta^{(2)}(\mathbf P_T+\mathbf l_T)
	\nonumber\\
	&\times
	F_c
	\sum_{i,j}
	\mathrm{Tr}
	\left[
	L_i^{\sigma\alpha}
	{\cal G}_{\rho\sigma\alpha\beta}
	\left(L_j^{\rho\beta}\right)^\dagger
	\right].
	\label{eq:CS-partonic-common}
\end{align}
Here \(\alpha_s=g_s^2/(4\pi)\), \(l\) is the momentum of the final-state gluon, and \(L_i\) denotes the amplitude structure associated with the corresponding cut diagram.
The color factor is
\begin{equation}
	F_c
	=
	\mathrm{Tr}(T^aT^b)\mathrm{Tr}(T^aT^b)
	=
	\frac{N_c^2-1}{4}.
	\label{eq:color-factor-common}
\end{equation}
The tensor \({\cal G}_{\rho\sigma\alpha\beta}\), which originates from the gluon field-strength tensors and the parent-gluon polarization average in the Collins--Soper definition, is defined as
\begin{equation}
	{\cal G}_{\rho\sigma\alpha\beta}
	=
	g_{\alpha\beta}g^{\mu\nu}
	\left(g_{\mu\rho}k^+ - k_\mu n_\rho\right)
	\left(g_{\nu\sigma}k^+ - k_\nu n_\sigma\right).
	\label{eq:Gtensor-common}
\end{equation}
The amplitude structures associated with the two ordinary QCD diagrams are
\begin{align}
	L_1^{\sigma\alpha}
	&=
	\frac{
		\mathrm{Tr}
		\left[
		\gamma^\sigma
		\left(
		\frac{\slashed P}{2}
		+\slashed q+\slashed l+m
		\right)
		\gamma^\alpha
		\Pi_1({}^1S_0^{[1]})
		\right]
	}{
		\left(\frac{P}{2}+q+l\right)^2-m^2
	},
	\nonumber\\
	L_2^{\rho\beta}
	&=
	\frac{
		\mathrm{Tr}
		\left[
		\gamma^\beta
		\left(
		-\frac{\slashed P}{2}
		+\slashed q-\slashed l+m
		\right)
		\gamma^\rho
		\Pi_1({}^1S_0^{[1]})
		\right]
	}{
		\left(\frac{P}{2}-q+l\right)^2-m^2
	}.
	\label{eq:L1L2-common}
\end{align}
Here \(\Pi_1({}^1S_0^{[1]})\) is the spin-singlet and color-singlet projection operator,
\begin{equation}
	\Pi_1({}^1S_0^{[1]})
	=
	\frac{
		(\slashed{\bar p}-m)\gamma_5(\slashed P+2E)(\slashed p+m)
	}{
		8\sqrt{2}E^2(E+m)
	}
	\otimes
	\frac{1_c}{\sqrt{N_c}} .
	\label{eq:projector-1S0-common}
\end{equation}
Here \(1_c/\sqrt{N_c}\) is the color projection operator for color-singlet states.

After using the delta functions to perform the phase-space integrations, we introduce the transverse-momentum variable
\(\rho=\mathbf P_T^2\).
The full QCD partonic fragmentation function can then be written as
\begin{equation}
	D_{g\to Q\bar Q({}^1S_0^{[1]})}(z,\mathbf q^2)
	=
	\alpha_s^2
	\int_0^\infty d\rho\,
	{\cal A}_{\rm fin}(z,\rho;\mathbf q^2).
	\label{eq:rho-integral-finite}
\end{equation}
Here
\({\cal A}_{\rm fin}(z,\rho;\mathbf q^2)\)
is the full QCD integrand that retains the complete dependence on the finite relative momentum.

For NRQCD matching, we divide the full QCD partonic fragmentation function by the leading perturbative LDME \(2N_c\) and define the finite-relative-momentum matching function as
\begin{equation}
	d_{\rm fin}(z,\mathbf q^2)
	=
	\frac{1}{2N_c}
	D_{g\to Q\bar Q({}^1S_0^{[1]})}(z,\mathbf q^2)
	=
	\frac{\alpha_s^2}{2N_c}
	\int_0^\infty d\rho\,
	{\cal A}_{\rm fin}(z,\rho;\mathbf q^2).
	\label{eq:d-fin-definition}
\end{equation}
Expanding \(d_{\rm fin}(z,\mathbf q^2)\) in powers of
\(\mathbf q^2/m^2\)
yields the fixed-order SDCs, whereas retaining its full finite-\(\mathbf q^2\) dependence provides the starting point for the resummation of relativistic corrections.

\section{Fixed-Order \(\mathcal O(v^4)\) Relativistic Corrections}
\label{sec:v4-correction}

In this section, we calculate the fixed-order relativistic corrections to gluon fragmentation into the color-singlet
\({}^1S_0^{[1]}\)
channel based on the Collins--Soper definition and the NRQCD partonic matching relation.
We expand the full QCD partonic fragmentation function in powers of
\(\mathbf q^2/m^2\)
and extract the SDCs
\(d^{(0)}(z)\),
\(d^{(2)}(z)\),
and the combined \(\mathcal O(v^4)\) SDC
\(d^{(4)}(z)\).

For a pseudoscalar heavy quarkonium \(H\), the NRQCD factorization formula for the color-singlet
\({}^1S_0^{[1]}\)
channel through \(\mathcal O(v^4)\) can be written as
\begin{align}
	D_{g\to H}^{\rm fix}(z)\big|_{v^4}
	=&\;
	d^{(0)}(z)\,
	\langle {\cal O}^{H}_1({}^1S_0)\rangle
	+
	d^{(2)}(z)\,
	\frac{\langle {\cal P}^{H}_1({}^1S_0)\rangle}{m^2}
	\nonumber\\
	&+
	d^{(4)}_1(z)\,
	\frac{\langle {\cal Q}^{H}_1({}^1S_0)\rangle}{m^4}
	+
	d^{(4)}_2(z)\,
	\frac{\langle {\cal Q}^{H}_2({}^1S_0)\rangle}{m^4}
	+
	d^{(4)}_3(z)\,
	\frac{\langle {\cal Q}^{H}_3({}^1S_0)\rangle}{m^4}
	+
	\mathcal O(v^6).
	\label{eq:NRQCD-factorization-v4}
\end{align}
Here
\(D_{g\to H}^{\rm fix}(z)\big|_{v^4}\)
denotes the fixed-order fragmentation function through
\(\mathcal O(v^4)\).
The quantities
\(d^{(0)}(z)\),
\(d^{(2)}(z)\),
and \(d^{(4)}_i(z)\)
are the SDCs associated with the corresponding NRQCD LDMEs.

The relevant NRQCD production operators are defined as
\begin{align}
	{\cal O}^{H}_1({}^1S_0)
	&=
	\sum_X
	\chi^\dagger\psi
	|H+X\rangle\langle H+X|
	\psi^\dagger\chi ,
	\nonumber\\
	{\cal P}^{H}_1({}^1S_0)
	&=
	\frac{1}{2}
	\sum_X
	\left[
	\chi^\dagger\psi
	|H+X\rangle\langle H+X|
	\psi^\dagger
	\left(-\frac{i}{2}\overleftrightarrow{\mathbf D}\right)^2
	\chi
	+\mathrm{H.c.}
	\right],
	\nonumber\\
	{\cal Q}^{H}_1({}^1S_0)
	&=
	\sum_X
	\chi^\dagger
	\left(-\frac{i}{2}\overleftrightarrow{\mathbf D}\right)^2
	\psi
	|H+X\rangle\langle H+X|
	\psi^\dagger
	\left(-\frac{i}{2}\overleftrightarrow{\mathbf D}\right)^2
	\chi ,
	\nonumber\\
	{\cal Q}^{H}_2({}^1S_0)
	&=
	\frac{1}{2}
	\sum_X
	\left[
	\chi^\dagger\psi
	|H+X\rangle\langle H+X|
	\psi^\dagger
	\left(-\frac{i}{2}\overleftrightarrow{\mathbf D}\right)^4
	\chi
	+\mathrm{H.c.}
	\right],
	\nonumber\\
	{\cal Q}^{H}_3({}^1S_0)
	&=
	\frac{1}{2}
	\left[
	\chi^\dagger\psi
	\sum_X |H+X\rangle\langle H+X|
	\psi^\dagger
	\left(
	\overleftrightarrow{\mathbf D}\cdot g_s\mathbf E
	+
	g_s\mathbf E\cdot\overleftrightarrow{\mathbf D}
	\right)
	\chi
	\right.
	\nonumber\\
	&~~~~~~\left.
	-
	\chi^\dagger
	\left(
	\overleftrightarrow{\mathbf D}\cdot g_s\mathbf E
	+
	g_s\mathbf E\cdot\overleftrightarrow{\mathbf D}
	\right)
	\psi
	\sum_X |H+X\rangle\langle H+X|
	\psi^\dagger\chi
	\right].
	\label{eq:NRQCD-operators-v4}
\end{align}
Here \(\psi\) and \(\chi\) are the heavy-quark and heavy-antiquark Pauli fields in NRQCD, respectively.
The derivative
\(\overleftrightarrow{\mathbf D}
=\overrightarrow{\mathbf D}-\overleftarrow{\mathbf D}\)
is the difference between the spatial covariant derivatives acting to the right and to the left.
The quantity \(\mathbf E\) is the chromoelectric field, and
\(\mathrm{H.c.}\) denotes the Hermitian conjugate.
The operators
\({\cal O}^{H}_1({}^1S_0)\)
and
\({\cal P}^{H}_1({}^1S_0)\)
correspond to the LO and \(\mathcal O(v^2)\) LDMEs, respectively.
The operators
\({\cal Q}^{H}_1({}^1S_0)\),
\({\cal Q}^{H}_2({}^1S_0)\),
and
\({\cal Q}^{H}_3({}^1S_0)\)
contribute at \(\mathcal O(v^4)\), with
\({\cal Q}^{H}_3({}^1S_0)\)
describing the interaction between the heavy-quark pair and the chromoelectric field.

For \({}^1S_0^{[1]}\)
heavy-quark pair, the perturbative matrix elements of
\({\cal Q}^{H}_1({}^1S_0)\)
and
\({\cal Q}^{H}_2({}^1S_0)\)
are both proportional to
\(2N_c\mathbf q^4\).
Consequently, matching with a free heavy-quark pair determines only the sum of
\(d^{(4)}_1(z)\)
and
\(d^{(4)}_2(z)\), rather than the two SDCs separately.
We therefore define the combined \(\mathcal O(v^4)\) SDC as
\begin{equation}
	d^{(4)}(z)
	=
	d^{(4)}_1(z)+d^{(4)}_2(z).
	\label{eq:d4-combination-v4}
\end{equation}
The chromoelectric operator
\({\cal Q}^{H}_3({}^1S_0)\)
involves the coupling of the heavy-quark pair to a soft gluon field.
Its SDC cannot be determined from the relative-momentum expansion of a free
\(Q\bar Q\)
state without external soft gluons and is therefore not included in
\(d^{(4)}(z)\).

Using Eqs.~\eqref{eq:kinematics-common} and
\eqref{eq:perturbative-LDME-general}, and applying
Eq.~\eqref{eq:NRQCD-factorization-v4}
to \({}^1S_0^{[1]}\) heavy-quark pair, we obtain the fixed-order partonic matching relation
\begin{equation}
	D_{g\to Q\bar Q({}^1S_0^{[1]})}^{\rm fix}
	(z,\mathbf q^2)\big|_{v^4}
	=
	2N_c
	\left[
	d^{(0)}(z)
	+
	\frac{\mathbf q^2}{m^2}d^{(2)}(z)
	+
	\frac{\mathbf q^4}{m^4}d^{(4)}(z)
	\right]
	+
	\mathcal O(v^6).
	\label{eq:partonic-expanded-v4}
\end{equation}
This equation relates the full QCD partonic fragmentation function to the fixed-order SDCs.

In the previous section, we defined the full QCD integrand
\({\cal A}_{\rm fin}(z,\rho;\mathbf q^2)\)
and the corresponding finite-relative-momentum matching function
\(d_{\rm fin}(z,\mathbf q^2)\).
To extract the fixed-order SDCs, we expand
\({\cal A}_{\rm fin}(z,\rho;\mathbf q^2)\)
about \(\mathbf q^2=0\),
\begin{equation}
	{\cal A}_{\rm fin}(z,\rho;\mathbf q^2)
	=
	{\cal A}_0(z,\rho)
	+
	\frac{\mathbf q^2}{m^2}{\cal A}_2(z,\rho)
	+
	\frac{\mathbf q^4}{m^4}{\cal A}_4(z,\rho)
	+
	\mathcal O(v^6).
	\label{eq:A-fin-expand-v4}
\end{equation}
Here
\({\cal A}_0(z,\rho)\),
\({\cal A}_2(z,\rho)\),
and
\({\cal A}_4(z,\rho)\)
are the coefficients at
\(\mathcal O(v^0)\),
\(\mathcal O(v^2)\),
and
\(\mathcal O(v^4)\), respectively.
Substituting Eq.~\eqref{eq:A-fin-expand-v4} into
Eq.~\eqref{eq:d-fin-definition}
and comparing with Eq.~\eqref{eq:partonic-expanded-v4}, we obtain
\begin{align}
	d^{(0)}(z)
	&=
	\frac{\alpha_s^2}{2N_c}
	\int_0^\infty d\rho\,
	{\cal A}_0(z,\rho),
	\nonumber\\
	d^{(2)}(z)
	&=
	\frac{\alpha_s^2}{2N_c}
	\int_0^\infty d\rho\,
	{\cal A}_2(z,\rho),
	\nonumber\\
	d^{(4)}(z)
	&=
	\frac{\alpha_s^2}{2N_c}
	\int_0^\infty d\rho\,
	{\cal A}_4(z,\rho).
	\label{eq:d024-from-A024-v4}
\end{align}
Eq.~\eqref{eq:d024-from-A024-v4} gives the direct relations for extracting the fixed-order SDCs from the expansion of the full QCD result.

We denote the projected full QCD amplitude by
\begin{equation}
	\mathcal M(q)
	\equiv
	\widetilde{\mathcal M}(q,E),
	\label{eq:M-tilde-definition-v4}
\end{equation}
where
\(\widetilde{\mathcal M}(q,E)\)
retains the dependence on the finite relative momentum and the heavy-quark energy.
The fixed-order expansion through
\(\mathcal O(v^4)\)
requires expanding both the explicit dependence on \(q\) and the implicit dependence through
\(E(\mathbf q^2)\),
\begin{equation}
	E
	=
	m+\frac{\mathbf q^2}{2m}
	-\frac{\mathbf q^4}{8m^3}
	+
	\mathcal O(v^6).
	\label{eq:E-expand-v4}
\end{equation}

After the \(S\)-wave angular average, the amplitude can be expanded as
\begin{equation}
	\mathcal M_{S\text{-wave}}
	=
	\mathcal M_0
	+
	\frac{\mathbf q^2}{m^2}\mathcal M_2
	+
	\frac{\mathbf q^4}{m^4}\mathcal M_4
	+
	\mathcal O(v^6),
	\label{eq:M-Swave-expand-v4}
\end{equation}
where
\begin{align}
	\mathcal M_0
	&=
	\left[\mathcal M(q)\right]_{q=0},
	\nonumber\\
	\mathcal M_2
	&=
	\frac{m^2}{2!(d-1)}
	I^{\alpha\beta}
	\left[
	\frac{\partial^2\mathcal M(q)}
	{\partial q^\alpha\partial q^\beta}
	\right]_{q=0},
	\nonumber\\
	\mathcal M_4
	&=
	\frac{m^4}{4!(d-1)(d+1)}
	I^{\alpha\beta\gamma\delta}
	\left[
	\frac{\partial^4\mathcal M(q)}
	{\partial q^\alpha\partial q^\beta
		\partial q^\gamma\partial q^\delta}
	\right]_{q=0}.
	\label{eq:M024-def-v4}
\end{align}
The transverse tensors are defined as
\begin{equation}
	I^{\alpha\beta}
	=
	-g^{\alpha\beta}
	+
	\frac{P^\alpha P^\beta}{4E^2},
	\qquad
	I^{\alpha\beta\gamma\delta}
	=
	I^{\alpha\beta}I^{\gamma\delta}
	+
	I^{\alpha\gamma}I^{\beta\delta}
	+
	I^{\alpha\delta}I^{\beta\gamma}.
	\label{eq:I-tensors-v4}
\end{equation}
Eq.~\eqref{eq:M024-def-v4} describes the \(S\)-wave expansion of the explicit relative-momentum dependence.
Since
\(\widetilde{\mathcal M}(q,E)\)
also depends implicitly on the relative momentum through
\(E(\mathbf q^2)\),
the \(\mathcal O(v^4)\) correction receives contributions from the explicit \(q\)-derivative terms, the energy-expansion terms, and their mixed terms.

Substituting Eq.~\eqref{eq:E-expand-v4} into
\(\widetilde{\mathcal M}(q,E)\),
the squared amplitude through
\(\mathcal O(v^4)\)
can be written as
\begin{align}
	|\mathcal M|^2
	=&
	\widetilde{\mathcal M}(0,m)
	\widetilde{\mathcal M}^{*}(0,m)
	\nonumber\\
	&+
	\frac{\mathbf q^2}{2!(d-1)}
	I^{\alpha\beta}
	\left[
	\frac{\partial^2\widetilde{\mathcal M}}
	{\partial q^\alpha\partial q^\beta}(0,m)
	\widetilde{\mathcal M}^{*}(0,m)
	+
	\mathrm{H.c.}
	\right]
	\nonumber\\
	&+
	\frac{\mathbf q^2}{2m}
	\left.
	\frac{\partial}{\partial E}
	\left[
	\widetilde{\mathcal M}(0,E)
	\widetilde{\mathcal M}^{*}(0,E)
	\right]
	\right|_{E=m}
	\nonumber\\
	&+
	\frac{\mathbf q^4}{[2!(d-1)]^2}
	I^{\alpha\beta}I^{\gamma\delta}
	\frac{\partial^2\widetilde{\mathcal M}}
	{\partial q^\alpha\partial q^\beta}(0,m)
	\frac{\partial^2\widetilde{\mathcal M}^{*}}
	{\partial q^\gamma\partial q^\delta}(0,m)
	\nonumber\\
	&+
	\frac{\mathbf q^4}{4!(d-1)(d+1)}
	I^{\alpha\beta\gamma\delta}
	\left[
	\frac{\partial^4\widetilde{\mathcal M}}
	{\partial q^\alpha\partial q^\beta
		\partial q^\gamma\partial q^\delta}(0,m)
	\widetilde{\mathcal M}^{*}(0,m)
	+
	\mathrm{H.c.}
	\right]
	\nonumber\\
	&+
	\frac{\mathbf q^4}{2m\,2!(d-1)}
	\left.
	\frac{\partial}{\partial E}
	\left\{
	I^{\alpha\beta}
	\left[
	\frac{\partial^2\widetilde{\mathcal M}}
	{\partial q^\alpha\partial q^\beta}(0,E)
	\widetilde{\mathcal M}^{*}(0,E)
	+
	\mathrm{H.c.}
	\right]
	\right\}
	\right|_{E=m}
	\nonumber\\
	&+
	\frac{\mathbf q^4}{8m^2}
	\left.
	\frac{\partial^2}{\partial E^2}
	\left[
	\widetilde{\mathcal M}(0,E)
	\widetilde{\mathcal M}^{*}(0,E)
	\right]
	\right|_{E=m}
	-
	\frac{\mathbf q^4}{8m^3}
	\left.
	\frac{\partial}{\partial E}
	\left[
	\widetilde{\mathcal M}(0,E)
	\widetilde{\mathcal M}^{*}(0,E)
	\right]
	\right|_{E=m}
	+
	\mathcal O(v^6).
	\label{eq:Msq-v4-expanded}
\end{align}

Substituting Eq.~\eqref{eq:Msq-v4-expanded} into the Collins--Soper cut-diagram expression in Eq.~\eqref{eq:CS-partonic-common} and performing the final-state phase-space integrations yield the expansion of the partonic fragmentation function in powers of
\(\mathbf q^2/m^2\).
The coefficients are identified as
\({\cal A}_0(z,\rho)\),
\({\cal A}_2(z,\rho)\),
and
\({\cal A}_4(z,\rho)\)
in Eq.~\eqref{eq:A-fin-expand-v4}.
After integrating over \(\rho\), comparison with the NRQCD partonic matching relation gives the fixed-order SDCs
\(d^{(0)}(z)\),
\(d^{(2)}(z)\),
and
\(d^{(4)}(z)\).

\section{Relativistic Resummation}
\label{sec:relativistic-resummation}

The fixed-order calculation extracts the SDCs
\(d^{(0)}(z)\), \(d^{(2)}(z)\), and \(d^{(4)}(z)\)
by expanding the finite-relative-momentum matching function
\(d_{\rm fin}(z,\mathbf q^2)\)
about \(\mathbf q^2=0\).
Truncating this expansion at \(\mathcal O(v^4)\) gives the fixed-order result through \(\mathcal O(v^4)\).
The resummation of relativistic corrections starts from the same
\(d_{\rm fin}(z,\mathbf q^2)\)
but retains its full finite-\(\mathbf q^2\) dependence
\cite{Gremm:1997dq,Bodwin:2007fz,Bodwin:2007ga,Fan:2012dy}.
Consequently, its low-order expansion reproduces the fixed-order SDCs, while the unexpanded expression includes a class of higher-order relativistic corrections generated by the same matching function.

To connect the finite-relative-momentum short-distance function with the NRQCD matrix elements of the physical quarkonium \(H\),
we introduce the relative-momentum moments
\begin{equation}
	\langle \mathbf q^{2r}\rangle_H
	=
	\frac{
		\langle {\cal O}^{H}_{1,2r}({}^1S_0)\rangle
	}{
		\langle {\cal O}^{H}_1({}^1S_0)\rangle
	},
	\qquad r=0,1,2,\ldots .
	\label{eq:q2r-ratio-resum}
\end{equation}
Here the numerator is the color-singlet LDME containing \(2r\) spatial covariant derivatives,
while the denominator is the LO color-singlet LDME.
Thus,
\(\langle \mathbf q^{2r}\rangle_H\)
represents the \(2r\)th relative-momentum moment inside the physical quarkonium.

Using the generalized Gremm--Kapustin relation, the higher-order relative-momentum moments can be approximated as
\cite{Gremm:1997dq,Bodwin:2007fz}
\begin{equation}
	\langle \mathbf q^{2r}\rangle_H
	\simeq
	\langle \mathbf q^2\rangle_H^r .
	\label{eq:GK-resum}
\end{equation}
This relation expresses the higher-order moments in terms of powers of
\(\langle \mathbf q^2\rangle_H\),
so that the finite-relative-momentum short-distance function
\(d_{\rm fin}(z,\mathbf q^2)\)
can be evaluated at
\(\mathbf q^2=\langle \mathbf q^2\rangle_H\).

The resummed gluon fragmentation function into \(H\) can then be written as
\begin{equation}
	D_{g\to H}^{\rm resum}(z)
	=
	\langle{\cal O}^{H}_1({}^1S_0)\rangle\,
	d_{\rm fin}
	\left(
	z,\langle \mathbf q^2\rangle_H
	\right).
	\label{eq:D-resum-final}
\end{equation}
Here
\(D_{g\to H}^{\rm resum}(z)\)
is the resummed fragmentation function, and
\(d_{\rm fin}(z,\mathbf q^2)\)
is the finite-relative-momentum matching function defined in
Eq.~\eqref{eq:d-fin-definition}.
Eq.~\eqref{eq:D-resum-final} evaluates the matching function at
\(\mathbf q^2=\langle \mathbf q^2\rangle_H\)
and therefore retains its full finite-\(\mathbf q^2\) dependence
\cite{Bodwin:2007fz,Bodwin:2007ga,Fan:2012dy}.

The correspondence between the resummed and fixed-order results follows directly by expanding
\(d_{\rm fin}(z,\langle \mathbf q^2\rangle_H)\)
in powers of
\(\langle \mathbf q^2\rangle_H/m^2\),
\begin{equation}
	d_{\rm fin}(z,\langle \mathbf q^2\rangle_H)
	=
	d^{(0)}(z)
	+
	\frac{\langle \mathbf q^2\rangle_H}{m^2}
	d^{(2)}(z)
	+
	\frac{\langle \mathbf q^2\rangle_H^2}{m^4}
	d^{(4)}(z)
	+
	\mathcal O(v^6).
	\label{eq:d-fin-resummed-expansion}
\end{equation}
Here
\(d^{(4)}(z)\)
is the combined \(\mathcal O(v^4)\) SDC defined in
Eq.~\eqref{eq:d4-combination-v4}.
Thus, the fixed-order result through \(\mathcal O(v^4)\) is recovered from the low-order expansion of the resummed expression, whereas
Eq.~\eqref{eq:D-resum-final}
further retains the relativistic corrections of \(\mathcal O(v^6)\) and higher generated by the same finite-relative-momentum matching function.

\section{Results and Discussion}
\label{sec:numerical-results}

In this section, we reproduce the LO result and the
\(\mathcal O(v^2)\)
relativistic correction
\cite{Braaten:1993rw,Gao:2016ihc}.
We then calculate the
\(\mathcal O(v^4)\)
relativistic correction and the resummed result for gluon fragmentation into the color-singlet and color-octet pseudoscalar heavy-quarkonium channels.
The fixed-order result through
\(\mathcal O(v^4)\)
is obtained from Eqs.~\eqref{eq:partonic-expanded-v4} and
\eqref{eq:d024-from-A024-v4}, whereas the resummed result is given by
Eq.~\eqref{eq:D-resum-final}.

\subsection*{A. Gluon fragmentation into \({}^1S_0^{[1]}\)}

For the fixed-order and resummed calculations, we use
\begin{equation}
	\langle v^4\rangle_H
	\simeq
	\langle v^2\rangle_H^2,
	\qquad
	\langle \mathbf q^2\rangle_H
	=
	m^2\langle v^2\rangle_H .
	\label{eq:v4-v2-q2-numerical}
\end{equation}
For the resummed result, the finite-relative-momentum matching function
\(d_{\rm fin}(z,\mathbf q^2)\)
is evaluated at
\(\mathbf q^2=\langle \mathbf q^2\rangle_H\).

The LO and \(\mathcal O(v^2)\) integrands are given in
Ref.~\cite{Gao:2016ihc}.
\begin{align}
	\mathcal{A}^{[1]}_0(z,\rho)
	=&
	\frac{
		2z(1-z)
		\left[
		\rho^2\bigl(z^2+(1-z)^2\bigr)
		+8\rho m^2(1-z)^3
		+16m^4(1-z)^4
		\right]
	}{
		mN_c
		\left[4m^2(1-z)+\rho\right]^2
		\left[4m^2(1-z)^2+\rho\right]^2
	}, \label{eq:A0-result}\\
	\mathcal{A}^{[1]}_2(z,\rho)
	=&
	-\frac{
		z(1-z)
	}{
		3N_cm
		\left[4m^2(1-z)+\rho\right]^3
		\left[4m^2(1-z)^2+\rho\right]^3
	}
	\nonumber\\
	&\times
	\Big[
	5\rho^4\bigl(z^2+(1-z)^2\bigr)
	-4\rho^3m^2(1-z)
	\left(34z^3-100z^2+81z-32\right)
	\nonumber\\
	&\quad
	+16\rho^2m^4(1-z)^3
	\left(80z^2-117z+66\right)
	-64\rho m^6(1-z)^5
	\left(51z-56\right)
	\nonumber\\
	&\quad
	+4352m^8(1-z)^7
	\Big].
	\label{eq:A2-result}
\end{align}
The \(\mathcal O(v^4)\) integrand is
\begin{equation}
	\mathcal A^{[1]}_4(z,\rho)
	=
	\frac{
		z(1-z)\,\mathcal N_4(z,\rho,m)
	}{
		180	N_cm
		\left[4m^2(1-z)+\rho\right]^4
		\left[4m^2(1-z)^2+\rho\right]^4
	},
	\label{eq:A4-result}
\end{equation}
here
\begin{align}
	\mathcal N_4(z,\rho,m)
	=&259 \rho ^6 \left(2 z^2-2 z+1\right)\nonumber\\
	&+8 m^2 \rho ^5 \left(1118 z^4-4431 z^3+6026 z^2-3790 z+1077\right)	\nonumber\\
	&+16 m^4 \rho ^4 (z-1)^2 \left(3878 z^4-20982 z^3+35358 z^2-24280 z+7965\right)\nonumber\\
	&-128 m^6 \rho ^3 (z-1)^4 \left(6837 z^3-20906 z^2+18860 z-7750\right)\nonumber\\
	&+256 m^8 \rho ^2 (z-1)^6 \left(16693 z^2-27740 z+16365\right)\nonumber\\
	&+7942144 m^{12} (z-1)^{10}-2048 m^{10} \rho  (z-1)^8 (3878 z-4437).
	\label{eq:N4-structure}
\end{align}
Eqs.~\eqref{eq:A0-result} and \eqref{eq:A2-result} agree with the results in Ref.~\cite{Gao:2016ihc}, while Eqs.~\eqref{eq:A4-result} and \eqref{eq:N4-structure} are new results obtained in this work.

\begin{figure}[htbp]
	\centering
	
	\begin{subfigure}[b]{0.32\textwidth}
		\centering
		\includegraphics[width=\textwidth]{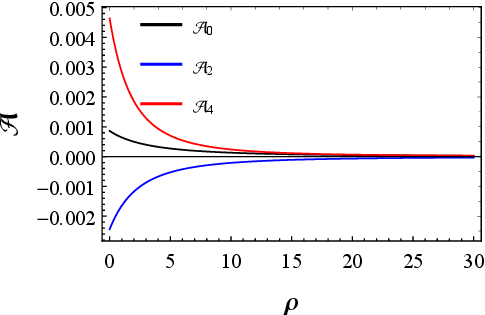}
		\caption{$z=0.1$}
		\label{fig:A_rho_z01}
	\end{subfigure}
	\hfill
	\begin{subfigure}[b]{0.32\textwidth}
		\centering
		\includegraphics[width=\textwidth]{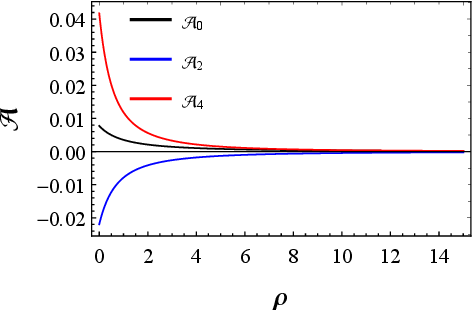}
		\caption{$z=0.5$}
		\label{fig:A_rho_z05}
	\end{subfigure}
	\hfill
	\begin{subfigure}[b]{0.32\textwidth}
		\centering
		\includegraphics[width=\textwidth]{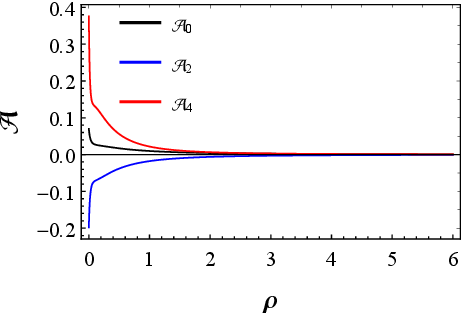}
		\caption{$z=0.9$}
		\label{fig:A_rho_z09}
	\end{subfigure}
\caption{
	(Color online) The fixed-order integrands
	\({\cal A}_0(z,\rho)\),
	\({\cal A}_2(z,\rho)\), and
	\({\cal A}_4(z,\rho)\)
	as functions of the transverse-momentum variable \(\rho\)
	for three representative values of the momentum fraction,
	\(z=0.1\), \(0.5\), and \(0.9\).
	The heavy-quark mass is set to
	\(m=1.4~{\rm GeV}\).
	In each panel, the three curves represent the LO integrand
	\({\cal A}_0\), the \(\mathcal O(v^2)\) integrand
	\({\cal A}_2\), and the \(\mathcal O(v^4)\) integrand
	\({\cal A}_4\), respectively.
}
	\label{fig:A_rho}
\end{figure}

We first examine the fixed-order integrands as functions of the transverse-momentum variable \(\rho\), as shown in Fig.~\ref{fig:A_rho}.
For a fixed momentum fraction \(z\), the integrands
\({\cal A}_0(z,\rho)\),
\({\cal A}_2(z,\rho)\), and
\({\cal A}_4(z,\rho)\)
are concentrated mainly in the small-\(\rho\) region and decrease rapidly as \(\rho\) increases.
This behavior shows that the \(\rho\) integration in the Collins--Soper definition is dominated by the small-transverse-momentum region, whereas the large-\(\rho\) contribution is strongly suppressed.
A comparison among different values of \(z\) shows that the integrands decrease more rapidly with \(\rho\) as \(z\) increases.
A larger value of \(z\) implies that the heavy-quark pair carries a larger fraction of the longitudinal momentum, leaving less phase space for the recoiling gluon.
Consequently, the large-transverse-momentum contribution is more strongly suppressed.
The large-\(z\) region is therefore concentrated at smaller transverse recoil, whereas the small-\(z\) region exhibits a broader distribution in \(\rho\).
The three integrands
\({\cal A}_0(z,\rho)\),
\({\cal A}_2(z,\rho)\), and
\({\cal A}_4(z,\rho)\)
also exhibit different dependences on \(\rho\).
Thus, the relativistic corrections are not proportional to the LO integrand point by point in \(\rho\).
Nevertheless, after integrating over \(\rho\) and performing the NRQCD matching, the
\(\mathcal O(v^2)\) and \(\mathcal O(v^4)\) SDCs are proportional to the LO SDC, with proportionality factors independent of \(z\).
This proportionality is therefore not a trivial property of the unintegrated expressions, but emerges only after the transverse-momentum integration.
The rapid decrease of all three integrands at large \(\rho\) ensures the convergence of the integral at its upper endpoint.
In addition, no nonintegrable singularity appears at the small-\(\rho\) endpoint, and the integrated SDCs contain no IR pole.
The
\(g\to Q\bar Q({}^1S_0^{[1]})\)
channel therefore remains IR finite at
\(\mathcal O(v^4)\).
This behavior differs from that of gluon fragmentation into the color-singlet
\({}^3S_1^{[1]}\)
channel, in which an IR divergence appears at the same order
\cite{Bodwin:2012xc}.

Substituting the fixed-order SDCs into the NRQCD factorization formula in
Eq.~\eqref{eq:NRQCD-factorization-v4}
and retaining terms through \(\mathcal O(v^4)\), we obtain the fixed-order fragmentation function
\begin{align}
	D_{g\to H}^{\rm fix}(z)\big|_{v^4}
	&=
	d^{(0)}(z)
	\langle \mathcal O_1^H({}^1S_0)\rangle
	\left[
	1-\frac{11}{6}\langle v^2\rangle_H
	+\frac{919}{360}\langle v^4\rangle_H
	\right]
	+O(v^6),\label{eq:DV4}\\
	d^{(0)}(z) &= \frac{\alpha_s^2}{4N_c^2 m^3} \left[ 3z - 2z^2 + 2(1 - z)\ln(1 - z) \right].
\end{align}
The resummed result differs from the fixed-order result through
\(\mathcal O(v^4)\)
in that the finite-relative-momentum matching function
\(d_{\rm fin}(z,\mathbf q^2)\)
is retained without truncating its expansion in
\(\mathbf q^2/m^2\)
and is evaluated directly at
\(\mathbf q^2=\langle \mathbf q^2\rangle_H\).
The corresponding resummed fragmentation function is
\begin{align}
	D_{g\to H}^{\rm resum}(z)
	=
	d^{(0)}(z)
	\langle \mathcal O_1^H({}^1S_0)\rangle
	\frac{ \ln^2 \left(\sqrt{1 + \langle v^2 \rangle} + \sqrt{\langle v^2 \rangle}\right)}{\langle v^2 \rangle (1 + \langle v^2 \rangle)^{3/2}}.
	\label{eq:Dresum}
\end{align}
Eq.~\eqref{eq:DV4} shows that the \(\mathcal O(v^2)\) correction is negative, whereas the \(\mathcal O(v^4)\) correction is positive.
Thus, the higher-order relativistic corrections to the
\(g\to {}^1S_0^{[1]}\)
channel do not simply continue the negative trend of the \(\mathcal O(v^2)\) correction.
Instead, the positive \(\mathcal O(v^4)\) contribution partially compensates for it.
Since the fixed-order SDCs
\(d^{(2)}(z)\)
and
\(d^{(4)}(z)\)
are both proportional to
\(d^{(0)}(z)\),
the fixed-order relativistic corrections do not alter the \(z\) dependence of the LO fragmentation function, but only modify its overall normalization.
By contrast, the resummed expression in Eq.~\eqref{eq:Dresum} retains the full finite-\(\mathbf q^2\) dependence of the finite-relative-momentum matching function and therefore includes a class of relativistic corrections of
\(\mathcal O(v^6)\),
\(\mathcal O(v^8)\),
and higher.
Expanding
\(D_{g\to H}^{\rm resum}(z)\)
in powers of
\(\langle\mathbf q^2\rangle_H/m^2\)
and using
\(\langle v^4\rangle_H\simeq\langle v^2\rangle_H^2\),
the terms through \(\mathcal O(v^4)\) reproduce
\(D_{g\to H}^{\rm fix}(z)\big|_{v^4}\).
The fixed-order result through \(\mathcal O(v^4)\) is therefore the corresponding truncation of the resummed expression, and the difference between the two represents the higher-order remainder generated by the same finite-relative-momentum matching function.

\begin{figure}[htbp]
	\centering
	
	\begin{subfigure}[b]{0.45\textwidth}
		\centering
		\includegraphics[width=\textwidth]{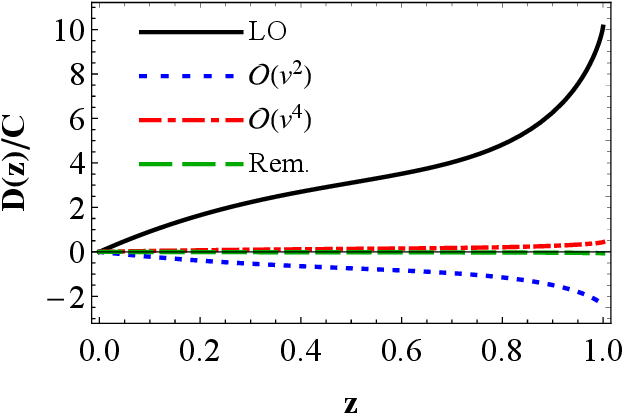}
		\caption{$g \to \eta_c + g$}
		\label{fig:fragmentation_function_mc}
	\end{subfigure}
	\hfill
	\begin{subfigure}[b]{0.45\textwidth}
		\centering
		\includegraphics[width=\textwidth]{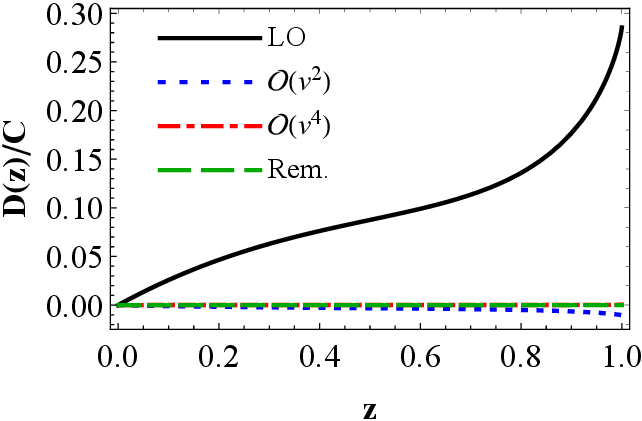}
		\caption{$g \to \eta_b + g$}
		\label{fig:fragmentation_function_mb}
	\end{subfigure}
\caption{
	(Color online) The normalized contributions to the gluon fragmentation function,
	\(D(z)/C\), as functions of the momentum fraction \(z\).
	The black solid, blue short-dashed, red dot-dashed, and green long-dashed curves
	denote the LO contribution, the \(\mathcal{O}(v^2)\) relativistic correction,
	the \(\mathcal{O}(v^4)\) relativistic correction, and the remaining higher-order contribution
	(denoted by Rem.), respectively.
	Here Rem. is obtained by subtracting the fixed-order result through
	\(\mathcal{O}(v^4)\) from the resummed result.
	In the numerical calculation, we take \(m_c=1.4~{\rm GeV}\) and
	\(\langle v_c^2\rangle=0.13\) for charmonium, and \(m_b=4.6~{\rm GeV}\) and
	\(\langle v_b^2\rangle=0.02\) for bottomonium.
	The normalization factor is chosen as
	\(C=10^{-3}\alpha_s^2\langle \mathcal{O}\rangle\).
}
	\label{fig:fragmentation_function}
\end{figure}

Fig.~\ref{fig:fragmentation_function} shows the effects of the fixed-order and resummed
relativistic corrections on the gluon fragmentation function.
The \(\mathcal O(v^2)\) correction is negative over the whole \(z\) range,
whereas the \(\mathcal O(v^4)\) correction is positive.
Therefore, the \(\mathcal O(v^4)\) contribution does not simply continue the negative trend of the
\(\mathcal O(v^2)\) correction, but partially compensates it.
A similar sign pattern was also found in the relativistic corrections to heavy-quark fragmentation
into color-singlet \(S\)-wave quarkonium \cite{Cui:2025wjq},
which indicates that the sign and numerical behavior of higher-order relativistic corrections
cannot be inferred directly from the lower-order result.
The remaining higher-order contribution represented by the green long-dashed curve is close to zero
over the whole \(z\) range.
This shows that, for the input parameters adopted in this work,
the \(\mathcal O(v^6)\) and higher-order relativistic corrections generated by the
finite-relative-momentum short-distance function \(d_{\rm fin}(z,\mathbf q^2)\) are numerically small.
Consequently, the fixed-order result through \(\mathcal O(v^4)\) already provides a good approximation
to the resummed result.
Comparing Fig.~\ref{fig:fragmentation_function_mc} with
Fig.~\ref{fig:fragmentation_function_mb}, one finds that the relativistic corrections are smaller
in bottomonium than in charmonium.
This behavior is consistent with the NRQCD velocity-scaling rules, since the typical heavy-quark
velocity is smaller in bottomonium and the higher-order terms in the velocity expansion are more
strongly suppressed.
Thus, the hierarchy of the velocity expansion is more pronounced for bottomonium.
Overall, the \(\mathcal O(v^4)\) correction preserves the hierarchy of the velocity expansion and
partially compensates the negative \(\mathcal O(v^2)\) correction.
The small remaining higher-order contribution further indicates that the velocity expansion
converges well for the input parameters considered here.

The ratio of each higher-order SDC to the LO SDC is defined as
\begin{equation}
	R_{v^{2r}}
	=
	\frac{d^{(2r)}(z)}{d^{(0)}(z)},
	\qquad r\geq 1 .
	\label{eq:R-v2n-def}
\end{equation}
This ratio quantifies the size and \(z\) dependence of each higher-order SDC relative to the LO SDC.
The corresponding contribution to the fragmentation function also depends on the associated velocity moment.
The fixed-order SDCs in Eq.~\eqref{eq:DV4} directly give
\(R_{v^2}\) and \(R_{v^4}\), with
\(R_{v^2}\) agreeing with the result of Ref.~\cite{Gao:2016ihc}.

The fixed-order result through \(\mathcal O(v^4)\) contains no relativistic corrections beyond
\(\mathcal O(v^4)\).
By contrast, expanding the finite-relative-momentum matching function entering
Eq.~\eqref{eq:Dresum}
in powers of \(\mathbf q^2/m^2\) yields
\(R_{v^2}\),
\(R_{v^4}\),
\(R_{v^6}(z)\),
\(R_{v^8}(z)\),
\(R_{v^{10}}(z)\),
and the corresponding higher-order ratios.
At \(\mathcal O(v^2)\) and \(\mathcal O(v^4)\), this expansion exactly reproduces the fixed-order SDC ratios.
The difference between the resummed result and the fixed-order result through
\(\mathcal O(v^4)\)
therefore arises entirely from the relativistic corrections of
\(\mathcal O(v^6)\) and higher generated by the same finite-relative-momentum matching function.
This agreement verifies the consistency between the resummed matching function and the fixed-order NRQCD matching results at low orders, while the higher-order ratios provide a direct means of examining the convergence of the velocity expansion.

Based on the relative sizes of the SDCs reflected by
\(R_{v^{2r}}(z)\), we further define the correction factor of the
physical fragmentation function relative to the LO result as
\begin{equation}
	K_X
	=
	\frac{
		D_{g\to H}^{X}(z)
	}{
		D_{g\to H}^{\rm LO}(z)
	},
	\qquad
	X=v^2,\,v^4,\,{\rm resum}.
	\label{eq:Kz-definition}
\end{equation}
Here \(K_X\) denotes the correction factor of the corresponding
fixed-order or resummed result relative to the LO fragmentation function.
We first obtain the fixed-order \(v^4\) results,
\begin{align}
	K_{v^4}
	&=
	1
	+
	R_{v^2}\langle v^2\rangle_H
	+
	R_{v^4}\langle v^4\rangle_H
	=
	1
	-\frac{11}{6}\langle v^2\rangle_H
	+
	\frac{919}{360}\langle v^4\rangle_H .
	\label{eq:K-v4-fixed}
\end{align}
We then obtain the resummed result,
\begin{equation}
	K_{\rm resum}
	=
	\frac{ \ln^2 \left(\sqrt{1 + \langle v^2 \rangle} + \sqrt{\langle v^2 \rangle}\right)}{\langle v^2 \rangle (1 + \langle v^2 \rangle)^{3/2}}.
	\label{eq:K-resum-def}
\end{equation}
The difference between the resummed result and the fixed-order \(v^4\) result is then given by
\begin{equation}
	K_{\rm resum}-K_{v^4}
	=
	-\frac{1073}{336}\langle v^2\rangle_H^3
	+
	\frac{761269}{201600}\langle v^2\rangle_H^4
	-\frac{57400241}{13305600}\langle v^2\rangle_H^5
	+\cdots .
	\label{eq:K-resum-minus-Kv4}
\end{equation}

\begin{figure}[htbp]
	\centering
	\includegraphics[width=0.45\textwidth]{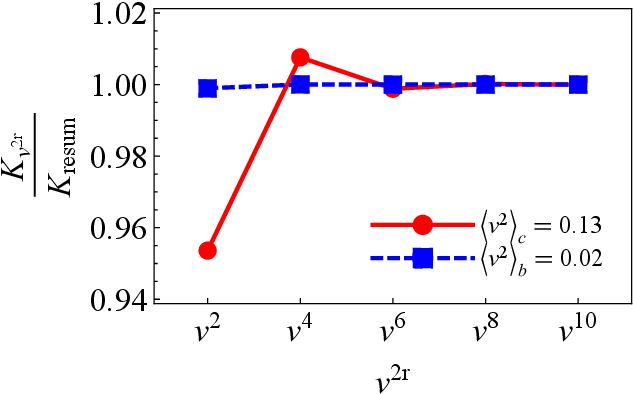}
	\caption{
		(Color online) The ratios of the cumulative fixed-order correction factors
		\(K_{v^{2r}}\) to the resummed correction factor
		\(K_{\rm resum}\), with \(r=1,2,3,4,5\).
		The red solid and blue dashed curves represent the charmonium result with
		\(\langle v_c^2\rangle=0.13\) and the bottomonium result with
		\(\langle v_b^2\rangle=0.02\), respectively.
	}
	\label{fig:ratio}
\end{figure}

According to the definition of \(K_X\) in Eq.~\eqref{eq:Kz-definition},
we compare the cumulative fixed order results through successive orders in the velocity expansion
with the resummed result using the ratios \(K_{v^{2r}}/K_{\rm resum}\),
as shown in Fig.~\ref{fig:ratio}.
As higher order relativistic corrections are successively included,
\(K_{v^{2r}}/K_{\rm resum}\) gradually approaches unity.
For charmonium, the fixed order \(v^4\) result differs from the resummed result by only \(0.76\%\).
For bottomonium, the deviation is further reduced to \(0.00258\%\).
This shows that the \(\mathcal O(v^6)\) and higher order relativistic corrections produce only very small additional effects.
Therefore, the velocity expansion converges well for both charmonium and bottomonium.
The convergence is faster for bottomonium because of its smaller value of \(\langle v_b^2\rangle\).

The average momentum fraction is
\begin{equation}
	\langle z\rangle_X
	=
	\frac{\int_0^1 dz\, z\,D_{g\to H}^{X}(z)}
	{\int_0^1 dz\,D_{g\to H}^{X}(z)}=	\frac{2}{3}.
	\label{eq:z-average-definition}
\end{equation}
Since \(\langle z\rangle_X\) is given by the ratio of the first moment to the zeroth moment,
it is insensitive to the overall normalization of the fragmentation function.
As in the fixed-order \(v^2\) result of Ref.~\cite{Gao:2016ihc},
the fixed-order \(v^4\) correction and the resummed relativistic correction calculated in this work
do not change the \(z\)-dependence of the fragmentation function,
but only modify its overall magnitude.
Therefore, the overall factor cancels between the numerator and the denominator in
Eq.~\eqref{eq:z-average-definition},
and \(\langle z\rangle_X\) takes the same constant value for
\(X=\mathrm{LO},\,v^2,\,v^4,\,\mathrm{resum}\).
This shows that the fixed-order and resummed relativistic corrections change the integrated fragmentation probability,
but do not change the average momentum fraction carried by the quarkonium.

\subsection*{B. Gluon fragmentation into \({}^1S_0^{[8]}\)}

For the tree-level relativistic corrections considered here, the color-octet and color-singlet channels have the same kinematic dependence and differ only by an overall color factor.
Accordingly, the fixed-order SDCs through \(\mathcal O(v^4)\) and the resummed SDCs satisfy
\begin{equation}
	d_{\rm fix}^{[8]}(z)\big|_{v^4}
	=
	{\cal R}_8\,
	d_{\rm fix}^{[1]}(z)\big|_{v^4},
	\qquad
	d_{\rm resum}^{[8]}(z)
	=
	{\cal R}_8\,
	d_{\rm resum}^{[1]}(z),
	\label{eq:octet-singlet-relation}
\end{equation}
where
\begin{equation}
	{\cal R}_8
	=
	\frac{N_c(N_c^2-4)}{N_c^2-1}.
	\label{eq:octet-color-factor}
\end{equation}
For \(N_c=3\), this factor becomes
$
	{\cal R}_8=\frac{15}{8}.
$
Since \({\cal R}_8\) is independent of \(z\), the color-octet and color-singlet channels have identical normalized relativistic corrections for both the fixed-order and resummed results.

\section{Summary}
\label{sec:summary}

Within NRQCD factorization and based on the Collins--Soper definition of fragmentation functions, we calculated the relativistic corrections to gluon fragmentation into pseudoscalar \(S\)-wave heavy-quark pairs in the
\({}^1S_0^{[1,8]}\)
channels.
By matching the full QCD result onto the NRQCD expression, we obtained the fixed-order SDCs through \(\mathcal O(v^4)\).
We also calculated the resummed relativistic correction using the finite-relative-momentum matching function
\(d_{\rm fin}(z,\mathbf q^2)\).
For the color-singlet
\({}^1S_0^{[1]}\)
channel, we reproduced the known \(\mathcal O(v^2)\) result and found that both the
\(\mathcal O(v^2)\) and \(\mathcal O(v^4)\) SDCs are proportional to the LO SDC.
Consequently, the fixed-order relativistic corrections modify only the overall normalization of the fragmentation function without changing its \(z\) dependence.
This channel remains IR finite at \(\mathcal O(v^4)\), in contrast to the color-singlet
\({}^3S_1^{[1]}\)
channel, in which an IR divergence appears at the same order.
Numerically, the negative \(\mathcal O(v^2)\) correction reduces the LO result, whereas the positive \(\mathcal O(v^4)\) correction partially compensates for it.
The fixed-order result through \(\mathcal O(v^4)\) is close to the resummed result, indicating that the relativistic corrections of \(\mathcal O(v^6)\) and higher are strongly suppressed and that the velocity expansion converges well for the input parameters considered here.
For the color-octet
\({}^1S_0^{[8]}\)
channel, the fixed-order and resummed SDCs differ from their color-singlet counterparts only by an overall color factor and therefore yield identical normalized relativistic corrections.
Our results provide the \(\mathcal O(v^4)\) short-distance inputs for gluon fragmentation into pseudoscalar \(S\)-wave quarkonia and can be applied to precision predictions for
\(\eta_c\) and \(\eta_b\)
production at large transverse momentum.

\section*{Acknowledgements} 
This work was supported by the National Natural Science Foundation of China (No. 11705078, 12575087).

\end{document}